\documentclass[conference]{IEEEtran}
\IEEEoverridecommandlockouts
\usepackage{cite}
\usepackage{amsmath,amssymb,amsfonts}
\usepackage{booktabs}
\usepackage{graphicx}
\usepackage{textcomp}
\usepackage{xcolor}
\usepackage{microtype}
\usepackage{url}

\def\BibTeX{{\rm B\kern-.05em{\sc i\kern-.025em b}\kern-.08em
    T\kern-.1667em\lower.7ex\hbox{E}\kern-.125emX}}

\begin{document}

\title{Green AI Carbon Optimizer: Carbon-Efficient Training Location
Recommendation and Global AI Energy Demand Forecasting}

\author{\IEEEauthorblockN{Yuxin Chen}
\IEEEauthorblockA{\textit{University of Helsinki, Finland} \\
yuxin.chen@helsinki.fi}
\and
\IEEEauthorblockN{Hao Gao}
\IEEEauthorblockA{\textit{Independent Researcher} \\
aoao4sci@gmail.com}
\and
\IEEEauthorblockN{Chujie Zou}
\IEEEauthorblockA{\textit{University of Helsinki, Finland}\\
chujie.zou@helsinki.fi}
}

\maketitle

\begin{abstract}
AI training and deployment consume substantial electricity, but carbon
outcomes remain weakly integrated into routine model development
decisions. This paper presents Green AI Carbon Optimizer with two
primary contributions: (i) a carbon aware cloud region recommendation
method for training workloads, and (ii) a power law forecasting pipeline
for global AI energy demand.

For location recommendation, we combine regional grid carbon intensity,
renewable share, and data center Power Usage Effectiveness (PUE) into a
unified scoring model across 100+ regions from major cloud providers.
For a reference workload (8$\times$A100, 100\,h), estimated emissions
in our sampled regions range from 7.74\,kg to 272.00\,kg CO$_2$.
Selecting the best region instead of the worst corresponds to a 97.2\%
reduction relative to the worst case. Ablation shows that ranking by
renewable share alone can select regions with higher CO$_2$ emissions
than rankings that include grid carbon intensity.

For forecasting, we fit a power law relation between parameter count and
training energy using 26 anchor models. We combine this fit with
scenario assumptions on model growth, hardware efficiency, and training
frequency, and evaluate sensitivity to inference ratio and ecosystem
scaling. Across scenarios, projected 2030 demand ranges from 7\,TWh to
1{,}436\,TWh under the stated assumptions, highlighting the importance
of deployment choices, model scaling discipline, and transparent energy
reporting.
\end{abstract}


\section{Introduction}

The scale of large language model training has made AI a visible source
of data center electricity demand. Training runs at frontier scale may
consume over 1{,}000\,MWh per model~\cite{b9}. However, carbon impact is
not determined by energy alone. For the same training workload, total
CO$_2$ can vary substantially across regions due to differences in grid
mix and facility efficiency.

Most practitioners still choose regions primarily by latency and price.
Existing tools such as CodeCarbon~\cite{b13} and MLCO2~\cite{b1} estimate emissions
for completed runs, but they do not provide a direct multi-cloud region
selection workflow before training begins. At the same time, macro-level
projections~\cite{b4,b5} are useful for policy context, but they do not
connect clearly to practical engineering decisions such as model size
planning, retraining frequency, or region choice.

This paper addresses both levels with one framework. The primary
contributions are:
\begin{enumerate}
  \item \textbf{Carbon-aware location scoring.} We propose a
  multi-factor region score using carbon intensity, renewable share, and
  PUE. The method supports cross-provider comparison and identifies large
  potential carbon reductions for fixed workloads.

  \item \textbf{Empirical scaling-law forecasting.} We fit
  $E = k\cdot P^\alpha$ on 26 anchor models and validate it against a
  Random Forest baseline under cross-validation for small-sample
  generalization.

  \item \textbf{Scenario and sensitivity analysis.} We project global AI
  energy demand to 2030 across four scenarios and quantify uncertainty
  from inference ratio, ecosystem multiplier, and PUE assumptions.
\end{enumerate}

Secondary contributions include an industry-trend extension model and a
personal carbon calculator module. We report these as exploratory and
application-oriented additions rather than core validated claims.

\section{Related Work}

Green AI work emphasizes the environmental externalities of
state-of-the-art model development~\cite{b6}. Reporting and estimation
tools~\cite{b1,b13} have improved visibility, but most remain post hoc.
Scaling-law literature~\cite{b3,b14} motivates functional forms that connect
model size and compute demand. Industry and policy reports~\cite{b4,b5}
provide macro-level electricity trajectories, while prior systems work
has examined carbon-aware datacenter computing and workload management
contexts~\cite{b7,b8}.

Our work differs in two ways. First, it provides an explicit
multi-provider location recommendation formulation tied to emissions.
Second, it links an empirical training-energy scaling law to
scenario-based demand projections with transparent assumptions.

\section{Methodology}

\subsection{Framework Overview}

The framework has two primary pipelines and two secondary modules.
Pipeline A computes region-level training emissions and ranks candidate
regions. Pipeline B forecasts aggregate AI energy through 2030 using an
empirical scaling law plus scenario assumptions.

Secondary module C introduces an exploratory industry-trend formulation
to capture organization growth, open-source retraining effects, and model
reuse assumptions. Secondary module D provides a user-facing personal
calculator that translates per-query energy into annual carbon impact.

\subsection{Dataset Construction}

The analysis uses three data layers.

\textbf{Layer 1: Anchor models.}
We compiled 26 large models with reported training energy or carbon,
spanning 65\,M to 2\,T parameters from public reports and
MLCO2-related sources~\cite{b1,b9}. Reported CO$_2$ values were
converted to kWh using 500\,g\,CO$_2$/kWh as a simplifying
global-average factor~\cite{b4}.

\textbf{Layer 2: Hugging Face measurements.}
We collected 75 Hugging Face model cards containing emissions-related
metadata, often linked to CodeCarbon-style reporting
workflows~\cite{b10,b15}. These data are used as a directional
cross-check, not as direct replacement targets for frontier pre-training
energy.

\textbf{Layer 3: Cloud infrastructure.}
We assembled carbon-intensity and renewable-share indicators for
100+ AWS, Azure, and GCP regions, together with regional PUE and GPU
availability proxies~\cite{b4,b11,b12a,b12b,b12c,b12d}. GPU
specifications were collected from official NVIDIA product pages.

\subsection{Carbon-Aware Location Scoring}

For workload configuration $(P_{\text{GPU}}, N_{\text{GPU}}, T)$, the
estimated training emissions in region $r$ are given by
\eqref{eq:carbon}:
\begin{equation}
  C_r = P_{\text{GPU}} \times N_{\text{GPU}} \times T \times I_r \times \mathrm{PUE}_r
  \label{eq:carbon}
\end{equation}
where $C_r$ is estimated training emissions (kg\,CO$_2$),
$P_{\text{GPU}}$ is average GPU power draw (kW),
$N_{\text{GPU}}$ is the number of GPUs, $T$ is training time (h),
$I_r$ is regional grid carbon intensity (kg\,CO$_2$/kWh), and
$\mathrm{PUE}_r$ is regional Power Usage Effectiveness (dimensionless).
For the reference workload used in Section~IV-B (A100 at 0.40\,kW,
8 GPUs, 100\,h), base IT energy is
$P_{\text{GPU}}N_{\text{GPU}}T = 320$\,kWh.

To rank regions consistently, we define the score in \eqref{eq:score}:
\begin{equation}
  S_r = 100\left[0.4(1-\hat{I}_r)+0.4R_r+0.2(2-\mathrm{PUE}_r)\right]
  \label{eq:score}
\end{equation}
where $\hat{I}_r \in [0,1]$ is normalized carbon intensity and
$R_r \in [0,1]$ is renewable share.
The score $S_r$ is a 0--100 ranking index (higher is better), and
$(2-\mathrm{PUE}_r)$ rewards lower PUE against reference upper bound 2.

We compare four ranking strategies in results: full score, carbon-only,
renewable-only, and carbon-only without PUE.

\subsection{Top-Down Forecasting Framework}

The top-down forecast has four steps.

\textbf{Step 1: Training-energy scaling law.}
We fit a log-linear model
\begin{equation}
  E = k \cdot P^\alpha,\quad k = 1.47\times10^{-4},\;\alpha=0.878
  \label{eq:scaling}
\end{equation}
on the 26 anchor models, with in-sample $R^2=0.854$.
Here, $P$ denotes absolute parameter count, $E$ is training energy
(kWh), and $(k,\alpha)$ are fitted coefficients.

\textbf{Step 2: Scenario settings.}
Each scenario is defined by annual parameter growth, annual hardware
efficiency improvement $r_{\text{hw}}$, and frontier training frequency.
We use:
\begin{itemize}
  \item Baseline: growth 100\%, $r_{\text{hw}}=10\%$, frequency 20/year.
  \item Moderate: growth 50\%, $r_{\text{hw}}=20\%$, frequency 12/year.
  \item High Efficiency: growth 30\%, $r_{\text{hw}}=30\%$, frequency 8/year.
  \item Green AI: growth 20\%, $r_{\text{hw}}=35\%$, frequency 5/year.
\end{itemize}
Hardware efficiency compounds as
$\eta(y)=(1-r_{\text{hw}})^{y-2024}$,
where $y$ is calendar year and $r_{\text{hw}}$ is annual hardware
efficiency improvement.

\textbf{Step 3: Energy accounting.}
We decompose total energy as in \eqref{eq:total}:
\begin{align}
  E_{\text{train-IT}} &= \text{GPU training energy} \notag \\
  E_{\text{IT}} &= E_{\text{train-IT}} \times (1+\rho_{\text{inf}}) \notag \\
  E_{\text{DC}} &= E_{\text{IT}} \times \overline{\mathrm{PUE}} \notag \\
  E_{\text{eco}} &= E_{\text{DC}} \times \mu_{\text{eco}}
  \label{eq:total}
\end{align}
where $E_{\text{train-IT}}$ is annual AI training IT energy,
$\rho_{\text{inf}}$ is the inference-to-training energy ratio,
$E_{\text{IT}}$ is total AI IT energy,
$\overline{\mathrm{PUE}}$ is average data-center PUE,
$E_{\text{DC}}$ is data-center energy after overhead, and
$\mu_{\text{eco}}$ is an ecosystem multiplier for upstream/downstream
energy not captured in direct IT load.
Defaults are $\rho_{\text{inf}}=50$, $\overline{\mathrm{PUE}}=1.3$, and
$\mu_{\text{eco}}=10$.

\textbf{Step 4: Carbon conversion.}
\begin{equation}
  \text{CO}_2(y) = E_{\text{eco}}(y)\times I_0
\end{equation}
where $I_0$ is average grid carbon intensity.
We use $I_0=500$\,g\,CO$_2$/kWh~\cite{b4,b11}; when reporting
Mt\,CO$_2$, we convert grams by $10^{12}$\,g/Mt.
Sensitivity ranges are reported in Section~\ref{sec:sensitivity}.

\subsection{Extended Industry-Trend Model and Personal Calculator}
\label{sec:extended}

This subsection is exploratory and intended to complement, not replace,
the top-down forecast.

For the industry-trend extension, we model aggregate demand through a
per-model term \eqref{eq:model_term} and an aggregate term
\eqref{eq:bottomup}:
\begin{equation}
  E_m(y)=\bar{P}e_0\left[(1-\gamma)+\gamma f_{\text{ft}}\right]\eta(y)\phi(y)
  \label{eq:model_term}
\end{equation}
\begin{equation}
  E_{\text{BU}}(y)=N_c(y)\cdot M_c(y)\cdot E_m(y)
  \label{eq:bottomup}
\end{equation}
where $y$ is year, $N_c(y)$ is the number of organizations training
frontier-class models in year $y$, $M_c(y)$ is average model releases per
organization, $\gamma$ is fine-tuning share, and $f_{\text{ft}}$ is
relative fine-tuning cost.
The term $E_m(y)$ denotes expected per-model training energy after
efficiency and open-source adjustments.
We use $\bar{P}$ in billion parameters per model and
$e_0$ in kWh per billion parameters, so $E_m(y)$ is in kWh/model and
$E_{\text{BU}}(y)$ is in kWh/year.
The efficiency factor $\eta(y)$ is the residual energy multiplier after
hardware improvement (values below 1 indicate lower energy).
In \eqref{eq:opensource}, the coefficient 2 is an exploratory assumption
for average open-source retraining intensity rather than an empirically
calibrated constant.

Open-source retraining is represented by \eqref{eq:opensource}:
\begin{equation}
  \phi(y)=1+2\,\omega(y)
  \label{eq:opensource}
\end{equation}
with $\omega(y)$ as the open-source retraining fraction in year $y$.

We also compute a sustainability indicator in \eqref{eq:sustainability}:
\begin{equation}
  \sigma(y)=\frac{E_{\text{BU}}(y)}{E_{\text{global}}(y)}\times100
  \label{eq:sustainability}
\end{equation}
where $\sigma(y)$ is the percentage share of global electricity and
$E_{\text{global}}(y)$ is projected global generation in year $y$.
The $E_{\text{global}}$ baseline follows IEA 2024 reporting~\cite{b4}.

For the personal module, daily energy is
$E_{\text{user}} = n_l e_l + n_m e_m + n_h e_h$,
where $n_l,n_m,n_h$ are daily light/medium/heavy query counts and
$e_l,e_m,e_h$ are per-query energy coefficients (Wh/query), so
$E_{\text{user}}$ is in Wh/day.
The per-query coefficients are compiled from published efficiency studies
and public model metadata~\cite{b2,b10}. Some values (e.g., GPT-5 and
Gemini~2.5~Pro) are extrapolated estimates based on scaling trends
rather than directly reported measurements. Carbon conversion uses
473\,g\,CO$_2$/kWh as a recent global average reference~\cite{b11}.

\section{Results}

\subsection{Scaling Law Validation}

The fitted law in \eqref{eq:scaling} captures 85.4\% of in-sample
variance and yields 5-fold CV $R^2=0.764\pm0.203$. Table
\ref{tab:model_compare} compares linear regression and Random Forest in
log-space prediction. Random Forest fits training data better, but it
underperforms in CV metrics on this small sample.

\begin{table}[t]
  \centering
  \caption{Model comparison on log-space energy prediction (5-fold CV,
  $n=26$). Units: log$_{10}$(kWh).}
  \label{tab:model_compare}
  \begin{tabular}{lccc}
    \toprule
    Model & CV $R^2$ & CV MAE & CV RMSE \\
    \midrule
    Linear Reg.\ (Scaling Law) & \textbf{0.764$\pm$0.203} & \textbf{0.317} & \textbf{0.421} \\
    Random Forest & 0.728$\pm$0.134 & 0.393 & 0.489 \\
    \bottomrule
  \end{tabular}
\end{table}

\subsection{Location Recommendation}

For the reference workload (8$\times$A100, 100\,h), the best-ranked
region in our dataset is Montr\'{e}al (GCP) at 7.74\,kg CO$_2$, while
Sydney (AWS) reaches 272.00\,kg. Table~\ref{tab:location} summarizes the
top and bottom cases computed from \eqref{eq:carbon}.

\begin{table}[t]
  \centering
  \caption{Top and bottom regions for the reference workload
  (8$\times$A100, 100\,h).}
  \label{tab:location}
  \begin{tabular}{llccc}
    \toprule
    Region & Provider & $I_r$ & PUE & CO$_2$ (kg) \\
    \midrule
    Montr\'{e}al & GCP & 0.022 & 1.10 & 7.74 \\
    Montr\'{e}al & AWS & 0.022 & 1.15 & 8.10 \\
    Stockholm & AWS & 0.040 & 1.10 & 14.08 \\
    S\~{a}o Paulo & GCP & 0.090 & 1.15 & 33.12 \\
    Oregon & GCP & 0.120 & 1.10 & 42.24 \\
    \midrule
    Singapore & AWS & 0.410 & 1.35 & 177.12 \\
    Tokyo & GCP & 0.510 & 1.13 & 184.42 \\
    Tokyo & AWS & 0.510 & 1.30 & 212.16 \\
    Sydney & GCP & 0.680 & 1.15 & 250.24 \\
    Sydney & AWS & 0.680 & 1.25 & 272.00 \\
    \bottomrule
  \end{tabular}
\end{table}

Ablation shows the effect of metric choice. Full score, carbon-only, and
carbon-only without PUE all select Montr\'{e}al (GCP) in this sample.
Renewable-only ranking selects Stockholm (AWS), which emits 14.08\,kg
for the same workload, 81.9\% higher than Montr\'{e}al. This indicates
that renewable share is not a sufficient proxy for operational emissions
without carbon intensity, which is why \eqref{eq:score} combines both.

\subsection{Top-Down Forecast (2025--2030)}

\begin{figure}[t]
  \centering
  \includegraphics[width=0.95\linewidth]{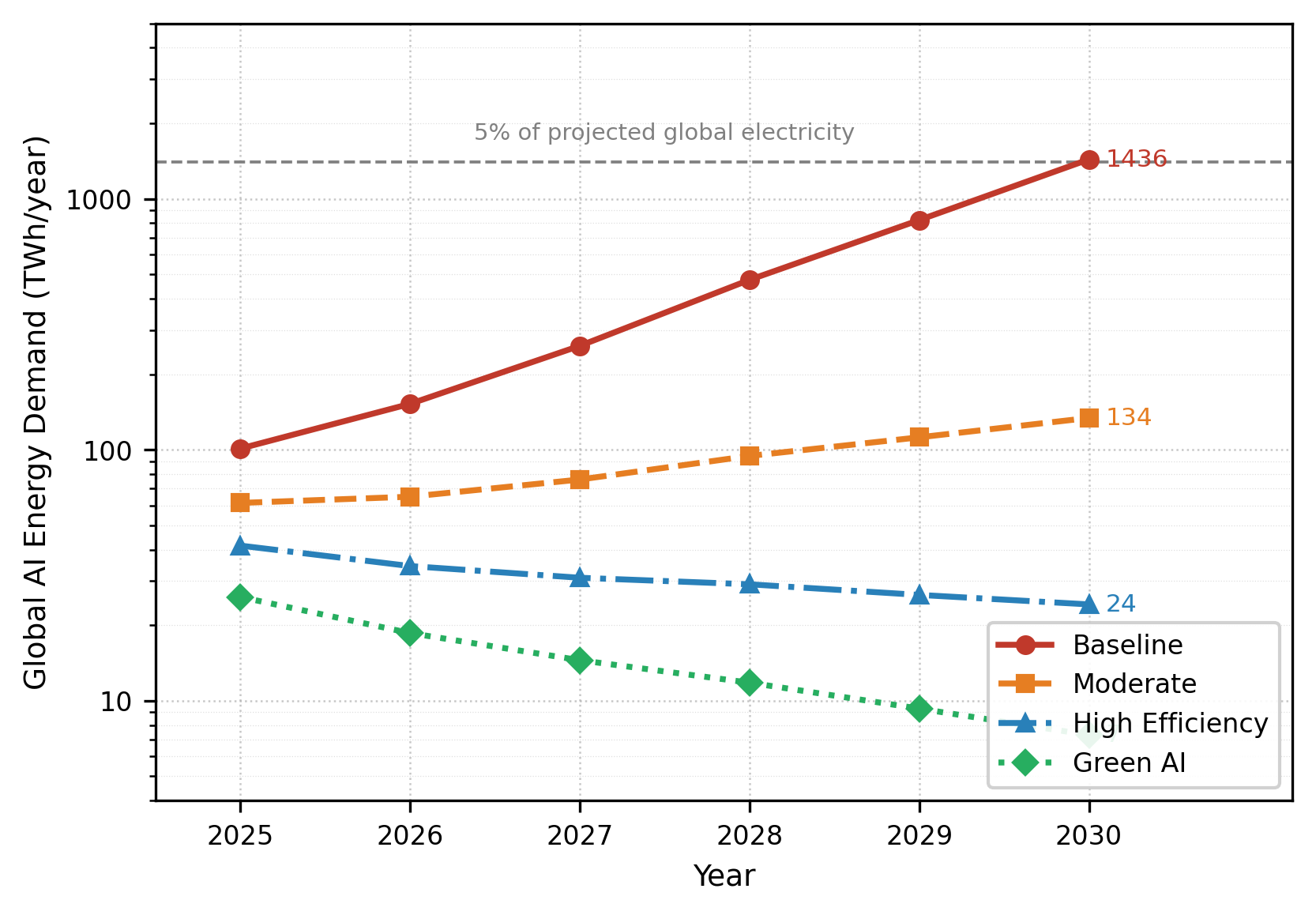}
  \caption{Projected global AI energy demand under four scenarios
  (log-scale $y$-axis). Parameters: $\rho_{\text{inf}}=50$,
  $\mu_{\text{eco}}=10$, $\overline{\mathrm{PUE}}=1.3$.}
  \label{fig:forecast}
\end{figure}

Table~\ref{tab:forecast} reports 2030 values under default parameters.
The Moderate scenario yields 134\,TWh/year, while Baseline reaches
1{,}436\,TWh/year. The Green AI scenario remains at 7\,TWh/year under
its stronger efficiency and lower-growth assumptions under
\eqref{eq:total}.

\begin{table}[t]
  \centering
  \caption{Top-down forecast: projected 2030 AI energy
  ($\rho_{\text{inf}}=50$, $\mu_{\text{eco}}=10$, $\overline{\mathrm{PUE}}=1.3$).}
  \label{tab:forecast}
  \begin{tabular}{lrrc}
    \toprule
    Scenario & Energy (TWh) & CO$_2$ (MT) & \% Global \\
    \midrule
    Baseline & 1{,}436 & 718 & 5.1\% \\
    Moderate & 134 & 67 & 0.5\% \\
    High Efficiency & 24 & 12 & 0.1\% \\
    Green AI & 7 & 4 & 0.03\% \\
    \bottomrule
  \end{tabular}
\end{table}

\subsection{Sensitivity Analysis}
\label{sec:sensitivity}

\begin{figure}[t]
  \centering
  \includegraphics[width=0.95\linewidth]{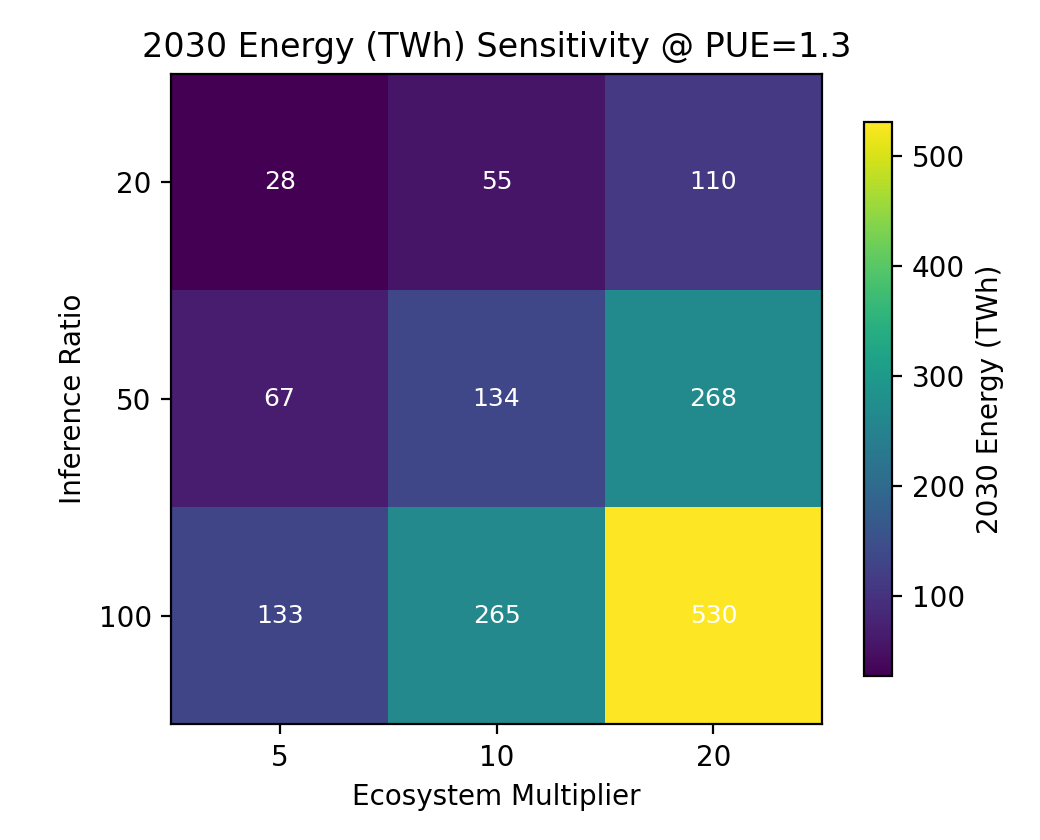}
  \caption{Sensitivity of 2030 global AI energy (TWh) to
  $\rho_{\text{inf}}$ and $\mu_{\text{eco}}$ at
  $\overline{\mathrm{PUE}}=1.3$ (Moderate scenario).}
  \label{fig:sensitivity}
\end{figure}

Under the Moderate scenario, the 2030 estimate spans from 27.6\,TWh
($\rho_{\text{inf}}=20$, $\mu_{\text{eco}}=5$) to 530.4\,TWh
($\rho_{\text{inf}}=100$, $\mu_{\text{eco}}=20$). The default Moderate
setting is 133.9\,TWh at $\rho_{\text{inf}}=50$, $\mu_{\text{eco}}=10$.
Changing $\overline{\mathrm{PUE}}$ from 1.1 to 1.6 has visible but smaller
impact than changing $\rho_{\text{inf}}$ or $\mu_{\text{eco}}$.

\subsection{Secondary Module Results}

Using the extended formulation in Section~\ref{sec:extended} and
\eqref{eq:bottomup}, the extension makes the growth mechanism explicit:
organization scale, model scale, hardware trend, and open-source
retraining jointly determine projected demand.
Under exploratory parameter settings, 2030
estimates range from approximately 1.4\,TWh (Sustainable) to
1{,}413\,TWh (Aggressive), with variation dominated by parameter
assumptions.
These outputs are used as scenario envelopes, not as calibrated
point forecasts.

With $\omega$ increasing from 55\% to 70\%, the retraining multiplier
$\phi(y)$ from \eqref{eq:opensource} rises from 2.10 to 2.40. Under the
same assumptions, increased
fine-tuning share reduces average per-model training demand, partially
offsetting growth in organization count and model-release activity.

Table~\ref{tab:perquery} reports per-query energy coefficients used in
the personal module. The spread across model families is substantial,
especially for medium and heavy workloads.

\begin{table}[t]
  \centering
  \caption{Per-query energy by model and complexity (Wh). Values are
  compiled from published efficiency studies~\cite{b2} and public model
  metadata~\cite{b10}; entries without direct public measurements
  (including GPT-5 and Gemini~2.5~Pro) are treated as order-of-magnitude
  extrapolated estimates.}
  \label{tab:perquery}
  \begin{tabular}{lccc}
    \toprule
    Model & Light & Medium & Heavy \\
    \midrule
    Gemini 2.5 Pro & 0.80 & 2.50 & 5.00 \\
    GPT-5 & 5.00 & 18.35 & 40.00 \\
    Claude 3.7 Sonnet & 0.84 & 2.00 & 5.52 \\
    DeepSeek-R1 & 23.80 & 30.00 & 33.60 \\
    \bottomrule
  \end{tabular}
\end{table}

For an example profile (5 light + 3 medium + 1 heavy queries/day on
GPT-5), daily use is about 120\,Wh, corresponding to roughly
21\,kg\,CO$_2$/year at 473\,g\,CO$_2$/kWh.

\section{Discussion}

\subsection{Empirical and Model-Driven Implications}

Two findings are robust across the presented evidence. First, location
choice can dominate training emissions for fixed workloads, with 97.2\%
variation between observed best and worst regions in our sample. Second,
the fitted scaling law captures broad energy trends with better
cross-validated behavior than a more flexible baseline on this
small-sample task.

The ablation analysis also shows that renewable share alone can be
misleading when carbon intensity differs substantially between regions.
This supports combined metrics for engineering decisions under
\eqref{eq:score} and \eqref{eq:carbon}.

The forecasting framework highlights where uncertainty enters. In our
sensitivity analysis, inference ratio and ecosystem multiplier dominate
variance, while PUE contributes less under the tested range. This means
single-point forecasts should be interpreted cautiously, and scenario
bands are more appropriate for decision support under \eqref{eq:total}.

The extended model makes the tradeoff between open-source retraining
growth and fine-tuning reuse explicit. These parameters are not
empirically calibrated in this work, so outputs from the extension are
best interpreted as structured hypotheses rather than validated
predictions despite the explicit structure in \eqref{eq:bottomup}.

\subsection{Practical and Policy Implications}

The location score and scaling-law estimator are intended for
design-phase planning. Given model size, hardware choice, and estimated
training duration, teams can compare carbon outcomes before committing to
a region. This complements post hoc measurement tools by adding an
upstream planning step.

The personal calculator is a communication module that translates model
selection and query behavior into user-scale energy terms. It is useful
for awareness and rough budgeting, but accuracy depends on model- and
deployment-specific telemetry quality.

For organizational reporting, the results suggest that average renewable
claims should not be used alone to infer operational emissions. Regional
carbon intensity and workload timing remain essential inputs.

At policy level, both forecasting frameworks indicate that high-growth
paths can approach material electricity shares by 2030 under stated
assumptions. This supports stronger disclosure standards for training and
inference energy and motivates evaluation of efficiency benchmarks beyond
accuracy alone.

\section{Threats to Validity}

This study uses public data with mixed reporting standards. Anchor-model
energy values are heterogeneous in provenance and may contain methodology
differences across sources. Regional carbon and PUE values are
time-varying and may not reflect short-term grid dynamics.

The scaling-law fit is based on 26 anchor models. While cross-validation
supports the chosen form for this sample size, the relation may shift
with future algorithmic and hardware changes. In the extended model and
personal calculator, several parameters are assumption-driven and should
be interpreted as approximate, so uncertainty is materially higher than
for the core location and top-down modules.

\section{Conclusion}

This paper presented Green AI Carbon Optimizer, a framework that links
carbon-aware region selection with scaling-law-based global demand
forecasting. The location module identifies large avoidable emissions for
fixed workloads, and the forecasting module provides a transparent
scenario engine with explicit sensitivity to deployment assumptions.
The strongest quantitative claims in this paper are grounded in the
location and top-down modules; the extension module is intended as a
structured exploratory mechanism.

Under default settings, projected 2030 AI electricity demand spans from
7\,TWh to 1{,}436\,TWh across scenarios. These results support three
practical directions: integrating carbon-aware placement into training
planning, improving reporting of inference energy, and calibrating future
forecasts with better public measurements of retraining and model reuse.

\end{document}